# Development and evaluation of an open-source, machine learning-based average annual daily traffic estimation software

**Zadid Khan***
**Ph.D. Student**
Glenn Department of Civil Engineering, Clemson University
351 Fluor Daniel Engineering Innovation Building, Clemson, SC 29634
Tel: (864) 359-7276, Fax: (864) 656-2670
Email: mdzadik@clemson.edu

**Sakib Mahmud Khan, Ph.D.**
Clemson University
Glenn Department of Civil Engineering
351 Fluor Daniel Engineering Innovation Building, Clemson, SC 29634
Tel: (864) 569-1082, Fax: (864) 656-2670
Email: sakibk@clemson.edu

**Mashrur Chowdhury, Ph.D., P.E., F.ASCE**
**Eugene Douglas Mays Endowed Professor of Transportation and**
**Professor of Automotive Engineering**
Clemson University
Glenn Department of Civil Engineering
216 Lowry Hall, Clemson, South Carolina 29634
Tel: (864) 656-3313   Fax: (864) 656-2670
Email: mac@clemson.edu

*Corresponding author

Abstract <u>223</u> + Text <u>4556</u> + 1 table: <u>250</u> + References <u>230</u> = **5229** words
Submission date: August 1, 2019

## ABSTRACT

Traditionally, Departments of Transportation (DOTs) use the factor-based model to estimate Annual Average Daily Traffic (AADT) from short-term traffic counts. The expansion factors, derived from the permanent traffic count stations, are applied to the short-term counts for AADT estimation. The inherent challenges of the factor-based method (i.e., grouping the count stations, applying proper expansion factors) make the estimated AADT values erroneous. Based on a survey conducted by the authors, 97% of the 39 public transportation agencies use the factor-based AADT estimation model, and these agencies face the aforementioned challenges while using factor-based models to estimate AADT. To derive a more accurate AADT, this paper presents the "estimAADTion" software, which is an open-source software developed based on a machine learning method called support vector regression (SVR) for estimating AADT using 24-hour short-term count data. DOTs conduct short-term counts at different locations periodically. This software has been designed to estimate AADT at a particular location from the short-term counts collected at those locations. In order to estimate AADT from short-term counts, the software uses data from permanent count stations to train the SVR model. The performance of the "estimAADTion" software is validated using the short-term count data from South Carolina. The Mean Absolute Percentage Error (MAPE) of the AADT estimated from the software is 3%, while the factor-based method produces a MAPE value of 6%.

## INTRODUCTION

Annual Average Daily Traffic (AADT) is defined as the average daily measure of the total volume of vehicles on a roadway segment over a year divided by 365 days. AADT is one of the most important traffic measures used in many transportation engineering projects (e.g., roadway design, transportation planning, and traffic safety analysis). In many cases, daily traffic volume can be directly counted each day with Automatic Traffic Recorders (ATR). However, installing and maintaining ATRs along every road is prohibitively expensive because of the cost associated with the installation and maintenance of ATRs on numerous roadways. To compensate for the lack of ATRs, AADT is often estimated using the short-term counts on an as-needed basis. Since AADT is needed for most transportation engineering projects, the accuracy of AADT estimation using short-term counts is critical for any transportation study that uses AADT as an input parameter.

For the locations without ATRs, traditionally AADT is estimated using the factor-based model where different expansion factors are used to estimate AADT from the short-term counts (Garber and Hoel, 2004). The expansion factors are estimated from the ATRs. These factors include growth and axle factor, and temporal (i.e., daily, monthly, seasonal) factors. Based on the geographic locations and roadway functional classes, the ATRs, and short-term count stations are grouped together to assign the expansion factors to the short-term count stations. However, state Departments of Transportation (DOTs) face challenges while using the factor-based method to estimate AADT using short-term counts. Due to the lack of established guidelines, the AADT estimation using a factor-based method often leads to less accurate AADT (Islam, 2016). Also having ATRs in the lower functional classes requires substantial financial investments. To overcome the challenges, an open-source, machine learning-based AADT estimation software is developed which can accurately estimate AADT using the 24-hour short-term counts for different roadway functional classes. Based on an earlier study conducted by authors, the machine learning-based method was found to estimate AADT more accurately compared to the factor-based method (Khan *et al.*, 2018). Following the earlier research, the 'estimAADTion' software is developed in this study to estimate AADT using the Support Vector Regression (SVR) model. The software can be used by any agencies for their jurisdiction, once the SVR model is trained with data from that corresponding jurisdiction. Also, this software enables agencies to estimate AADT using historic short-term count data using growth factors. The software is also able to collect and store ATR data from the website of any state DoT for future use. This software requires 24-hour counts as input. The software also aggregates the results from multiple 24 hour counts and provides only one AADT output for each ATR.

The goal of this study is to develop and evaluate an AADT estimation software and compare the performance of the model with the factor-based model. This paper discusses the existing practices of AADT estimation adopted by public transportation agencies to characterize the need for better AADT estimation models based on the findings from an online survey. Later, the estimAADTion software functions, user interface, and input-output files are discussed. A case study is conducted to evaluate the performance of the software using data from different roadway functional classes in South Carolina. The model performance is validated with new short-term count data that was not used for training. The software estimation (based on SVR model) is compared with factor-based estimation in terms of Mean Absolute Percentage Error (MAPE) for interstates, arterials, and collectors. The software is available in an online repository that can be accessed via "*https://github.com/zadid56/C-/tree/master/AADT/AADT*" link. The source code is open so anyone can download the software and use it in its current condition or update the source code to enhance the capabilities of the software.

## RELATED STUDIES ON AADT ESTIMATION

This section describes previous works related to AADT estimation. One of the most popular models for AADT estimation is the regression-based model. Different variations of regression models have been applied in different studies. These models include the Ordinary Least Squares (OLS) regression and the Geographically Weighted Regression (Mohamad *et al.*, 1998; Sharma *et al.*, 2001). Another AADT estimation model is the origin-destination centrality model which reduces the Root Mean Square Error to half compared to the travel demand-based AADT model (Chowdhury *et al.*, 2019).

Among the machine learning techniques, Artificial Neural Networks (ANNs) have also been used to determine AADT using short-term traffic counts (Sharma *et al.*, 1999). SVR is another popular method used for estimating AADT. A study by Lin indicated that, compared to ANN, SVR has greater learning potential (Lin, 2004). In another study, Castro-Neto et al. evaluated the performance of a modified SVR, which is SVR with data-dependent parameters (SVR-DP), in which the authors collected AADT values from 1985 and 2004 from Tennessee (Castro-Neto *et al.*, 2009). The authors found SVR model performs better than the popular Holt-Winters exponential smoothing (HW method) and the OLS linear regression methods.

## CURRENT PRACTICES OF AADT ESTIMATION BY PUBLIC TRANSPORTATION AGENCIES

An online survey was sent to U.S. states and Canadian province agencies to determine the standard practices of the agencies to estimate AADT. Thirty-eight states participated in the online survey, along with three Canadian provinces. When asked what method the agency uses for estimating AADT from short-term traffic count stations, 38 (97%) of 39 said the factor-based method, while only 4 said the regression method. 39 transportation agencies specified what conversion factors are used to estimate AADT from traffic volume data at short-term traffic count locations. 32 agencies said axle adjustment factors were used, while 30 said monthly adjustment factors were used to estimate AADT from short-term counts. 29 agencies said daily factors and 26 agencies said seasonal factors are used to derive AADT from short-term counts. The most commonly stated challenge in estimating AADT at short-term traffic count locations using the factor approach was maintaining enough permanent traffic count locations to ensure accurate factors, which was said by 28 of 40 respondents. Assigning individual short-term count locations to the correct permanent traffic count station group was said by 22 agencies and grouping the permanent count stations appropriately was said by 20 agencies as the AADT estimation challenges. 17 agencies said those grouping short-term traffic count stations was a challenge, while 11 said assigning the group of short-term count locations to the correct permanent traffic count station group was a challenge. The motivation of developing the estimAADTion software is to overcome the challenges that agencies face while using the factor-based AADT estimation model

The transportation agencies were then asked what revisions they currently plan to implement for data collection, processing, and estimation. Of the 40 agencies to respond, 22 said to add more permanent traffic count stations, 16 said to add more short-term traffic count locations, 12 said to re-group the permanent traffic count stations, 6 said they do not have any plans for further implementation. Using the estimAADTion software, the agencies can directly use short-term counts without worrying about the grouping of the stations.

Of 41 agencies to respond, 17 said that they use alternative methods to predict local roadway AADT. 24 agencies said that they do not use alternate methods. Alternate methods listed by the 17 agencies included historical counts by 7 agencies, and travel demand model data by 5 agencies to predict local roadway AADT. The estimAADTion software will enable agencies to compare their existing method with the new machine learning-based method. If adopted, the estimAADTion software can provide substantial resource savings for numerous transportation-related projects by estimating accurate AADT for different roadway functional classes.

## AADT ESTIMATION MODEL DEVELOPMENT

The most important aspect of the software is the AADT estimation model that runs in the background. This is the backbone of the software that enables accurate AADT estimation from the software. SVR model has been used as the AADT estimation model of the software. SVR is used for nonlinear regression by mapping the training dataset onto a higher dimensional, kernel-induced feature space. In this study, an SVR algorithm with a radial basis kernel function is chosen and implemented using the MATLAB LIBSVM library tool (Chang and Lin, 2011). The model contains several parameters such as margin ($\varepsilon$), cost function (C), and radial kernel parameter ($\gamma$). The $\varepsilon$ (epsilon) parameter controls the width of the $\varepsilon$ insensitive zone and is used to fit the training data. The cost function (C) is used to determine the tradeoff between the complexity of the model and the degree of deviation from $\varepsilon$ that can be tolerated. The initial value of C and $\gamma$ (gamma) are determined based on the grid search method with 5-fold cross-validation. Cross-validation is performed to reduce the bias of a training dataset on the model parameters. The trial and error method is used to find the parameters that yield the lowest mean-squared error in AADT prediction. C and $\gamma$ values are searched as powers of two, where the powers for C value vary within the − 3 to 15 range, and the powers for $\gamma$ value vary within the −15 to 3 range (Chang and Lin, 2011). The value of the parameters varied from model to model with the change in training data.

## estiMAADTion SOFTWARE FUNCTIONS

The functions of the estimAADTion software can be classified into two parts. In the first part, the software is capable of automatically collecting data from an online repository. Although the software is currently developed to collect data from the South Carolina Department of Transportation (SCDOT) website, it can be modified based on each agency and their data source, as the source code of the software is available in the github repository mentioned in the introduction section. In the second part, the software performs the AADT estimation based on the 24-hour short-term traffic count.

The estimAADTion software functions are described below.

## ATR Data Collection

In order to perform AADT estimation, the user needs to input the ATR data (i.e., hourly volume data from all permanent count stations) to the model. The software provides the capability of ATR data collection also. For this function, the user must also input several data to the software. The inputs are:

- List of ATRs
- Data Collection location
- Year

A script has been developed using Python programming language to automatically collect all the ATR data from any online repository. In this study, data has been collected from the SCDOT website (*http://dbw.scdot.org/Poll5WebAppPublic/wfrm/wfrmViewDataNightly.aspx*). The script will be provided to the users with the software package. The script collects the hourly volumes for each ATR in separate text files, which can be directly used by the AADT estimation part of the software. The user will need to specify the year of data collection and the folder where all the data will be saved. Although the primary use of the software is to estimate AADT of the current year, the year has been kept as an input so that users can choose to estimate AADT for a previous year also, if required. This process takes approximately a couple of hours depending on data availability, connectivity and hardware specifications of the computer. However, in an ideal situation, this procedure will only need to be done once per year. After downloading the ATR data, the user can reuse the data for the rest of the year, as often as necessary.

## AADT Estimation

AADT Estimation is the primary function of the software. The SVR model uses the open-source LIBSVM library. The SVR model uses the ATR data of a specific year to train the model, determines the model parameters, and estimates the AADT using short-term counts as inputs.

The user needs to provide several inputs to the software. The required inputs are:

- List of ATRs and their functional classes
- Short-term counts
- ATR data of that year (If new data is available for updating the model)
- Expansion factors for factor-based method
- SVR Model parameters (if the SVR model already trained with existing ATR data)
- Growth factor (for AADT estimation from the historic short-term count)

As the software estimates AADT based on the short-term count, the file which has the short term count data should be provided as an input file. The list of ATRs (with the information of the functional classes) and ATR data are required to train the SVR model. The SVR model parameters will be generated and saved in a file by the software when it will train the SVR model with new ATR data. If the same ATR data is used again, then the file with the SVR parameters can be entered as input so the model does not need to re-train with the data. The expansion factors are needed for the factor-based model prediction, and these factors can be obtained from the state transportation agencies for any specific year. The final input file is the file with the growth factor, which enables the user to use any historic short-term count data to estimate current year AADT. Using the growth factor, the short-term count for the current year is projected from the historic count, and then this data can be used as input to the software to estimate AADT.

## USER INTERFACE, INPUT FILES AND OUTPUT FILES FOR AADT ESTIMATION

The hardware and software requirements of the estimAADTion software and the process of input file preparations are available in the report published by the authors (Chowdhury *et al.*, 2019). The user interface and both input and output files are discussed in the following subsections.

### User Interface

The user interface of the software has three separate tabs for different types of inputs. The three tabs are described briefly below.

#### *AADT Estimation*

The first tab at the top of the window is designed for inputs related to the AADT estimation (Figure 1(a)).

From Figure 1(a), it can be observed that there are four 'browse' buttons, corresponding to four different input files. The user can browse to the location of the input files or copy-paste the locations of the input files as text in the white empty text boxes. The files must be in Comma Separated Value (.CSV) format. After providing all the inputs, the user can click the "Predict AADT" button at the bottom and the software will run the program. While the program runs in the background, a progress bar will show the progress of the software. After the run is completed, the software will provide the predicted AADT output in an excel file with the date and time at which the results are generated.

#### *ATR Data Collection*

The second tab at the top of the window is designed for inputs related to ATR data collection (Figure 1(b)).

From Figure 1(b), it can be observed that there are two 'browse' buttons, corresponding to two different input files. The first input is the location of the data collector script. The script will be provided to the user with the software package. The second input is the location of the data collection folder. The third input is not a location so it does not contain a browse button, it is the year of ATR data collection, in the format YYYY (i.e., 2017, 2018). After providing all the inputs, the user can click "Get Data" at the bottom and the software will run the program in the background. The data collection can be computationally intensive. The data collector script is written for a specific website from which data can be collected. If the website address changes and/or the actions for downloading the data changes, then the script needs to be modified by the user accordingly. The details are provided in the final report of this project (Chowdhury *et al.*, 2019).

#### *ATR List*

Regardless of the purpose of use, the software requires the list of ATRs to function, so a separate third tab has been created for this input. ATRs can be added or removed for various reasons. So, the ATR list needs to be updated accordingly in a CSV file and it should be uploaded whenever the software is used. Figure 1(c) shows the third tab of the software. The browse button can be used to browse the location of the ATR list file. This file has to be in CSV format. The user also has the option to copy and paste the location of the file in the blank text box.

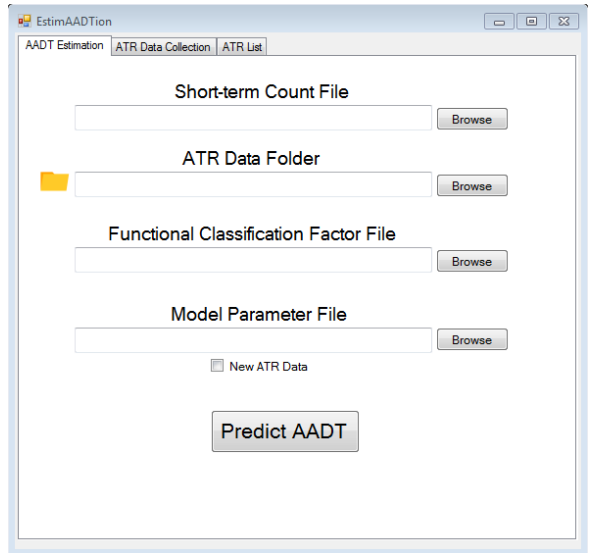

(a) AADT estimation Tab

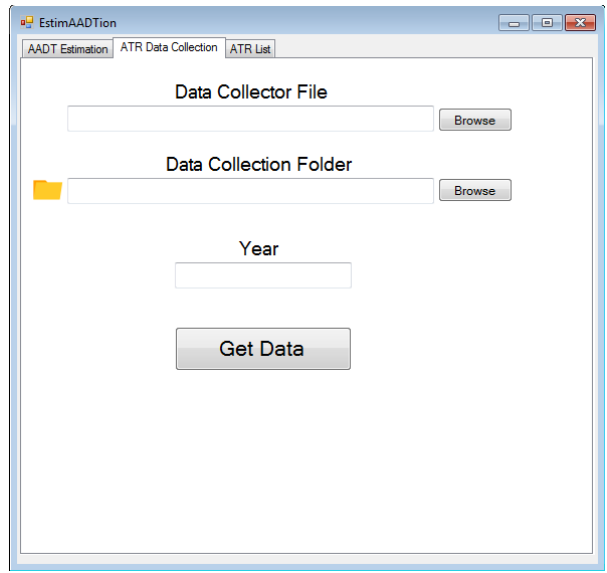

(b) ATR data collection tab

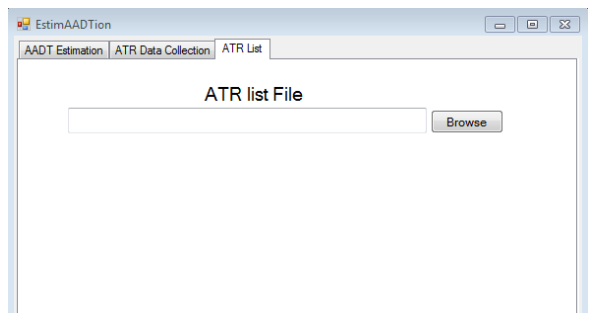

(c) ATR List Tab

**FIGURE 1 estimAADTion Software GUI**

**Input Files for AADT Estimation**

The previous subsection discussed about ATR data collection so it will not be discussed here. Other than ATR data, all other inputs are in the form of excel files (*.xlsx or *.csv). Since the input files need to follow a specific structure. The input file formats are described below.

*Short-term Count File*

This CSV file contains the short-term counts for AADT prediction (Figure 2(a)). Row 1 must contain the headings for each column. Column A contains the county and Column B contains the station number/ID. This information is provided to keep the location information. Column C contains the date of the short-term count. The date is used to identify the day of the week and month of the year, which influence the hourly volumes. Column D contains the functional class of the road at which the short-term count was taken. This information is used to determine which model should be used for AADT prediction. Column E contains the growth factor. The user can input an old short-term count and specify a growth factor. The software will use the growth factor to project the hourly volume to the current year. For all other short-term counts that are up-to-date, the value should be 1 in this column. A blank column is equivalent to 0, so inputting the value of 1 is mandatory. Starting in column F and for each subsequent column to its right, each column contains one-hour volume for each of the 24 hourly volumes. The software can aggregate the results from multiple 24 hour counts and provide only one AADT output for each ATR.

*Expansion Factor File*

The expansion factor CSV file contains the functional cla ssification factors (Figure 2(b)). Column A contains the labels. Row 1 contains all possible functional classes. There are 10 possible functional classes, which are placed in Columns B through K. Row 2 contains the Axle factors by functional class. Rows 3 through 14 contain the seasonal factors by functional class. As we know, the seasonal factors are on a monthly basis and there are 12 months. So, Row 3 contains the seasonal factors for the month of January, Row 4 contains the seasonal factors for the month of February. This trend continues to Row 14. If a cell is left blank, it is interpreted as 0 by the software.

*SVR Model Parameter File*

The Model Parameter CSV file is generated by the software when new ATR data is added. This CSV file will remain unchanged as long as there is no update to the ATR data (Figure 2(c)). The model has two primary parameters, C and $\gamma$ (gamma). Each row (2-4) correspond to one of the three prediction models in the software, interstate model, arterial model and collector model. The software will create a new model parameter file if the user checks the box "New ATR Data". This means that the software will re-train the SVR model and generate new parameters and save it in the CSV file specified by the user. This step should only be performed when new ATR Data arrives (ideally at the start of each year). The next time the user wants to use the software with the same ATR data, the user can just select the generated file and leave the "New ATR Data" box unchecked.

|   | A | B | C | D | E | F | G | H | I | J |
|---|---|---|---|---|---|---|---|---|---|---|
| 1 | County | Station | Date | FClass | GF | Hour1 | Hour2 | Hour3 | Hour4 | Hour5 |
| 2 | 1 | 80 | 10/19/2016 | 12 | 1 | 27 | 24 | 20 | 43 | 49 |
| 3 | 1 | 80 | 10/20/2016 | 12 | 2 | 26 | 27 | 20 | 20 | 40 |
| 4 | 1 | 99 | 10/19/2016 | 12 | 1 | 436 | 317 | 231 | 301 | 545 |
| 5 | 1 | 99 | 10/20/2016 | 12 | 1 | 459 | 293 | 285 | 318 | 547 |
| 6 | 1 | 9 | 10/24/2017 | 2 | 1 | 35 | 22 | 21 | 33 | 74 |
| 7 | 1 | 10 | 11/28/2017 | 4 | 1 | 6 | 9 | 2 | 6 | 7 |
| 8 | 1 | 11 | 11/14/2017 | 2 | 1 | 83 | 38 | 42 | 50 | 107 |
| 9 | 1 | 17 | 11/7/2017 | 4 | 1 | 9 | 6 | 9 | 12 | 12 |
| 10 | 1 | 65 | 10/24/2017 | 4 | 1 | 2 | 3 | 10 | 12 | 7 |
| 11 | 1 | 73 | 10/24/2017 | 4 | 1 | 26 | 7 | 10 | 16 | 28 |
| 12 | 1 | 133 | 11/7/2017 | 2 | 1 | 16 | 16 | 8 | 34 | 44 |
| 13 | 1 | 149 | 10/25/2017 | 2 | 1 | 138 | 74 | 76 | 84 | 212 |
| 14 | 1 | 151 | 11/7/2017 | 4 | 1 | 15 | 7 | 8 | 14 | 17 |
| 15 | 1 | 152 | 11/7/2017 | 4 | 1 | 8 | 18 | 6 | 5 | 21 |
| 16 | 1 | 153 | 11/7/2017 | 4 | 1 | 15 | 11 | 7 | 15 | 30 |
| 17 | 1 | 155 | 11/7/2017 | 4 | 1 | 43 | 25 | 21 | 20 | 54 |
| 18 | 1 | 156 | 11/7/2017 | 4 | 1 | 20 | 19 | 14 | 23 | 55 |

(a) Short-term count CSV file format

|   | A | B | C | D | E | F | G | H | I | J | K | L |
|---|---|---|---|---|---|---|---|---|---|---|---|---|
| 1 | FC | 2 | 3 | 4 | 5 | 9 | 12 | 13 | 14 | 15 | 18 | |
| 2 | Axle_f | 0.91 | 0.93 | 0.94 | 0.96 | 0.96 | 0.96 | 0.93 | 0.94 | 0.97 | 0.97 | |
| 3 | Seasonal_f | 1.15 | 1.07 | 1.05 | 1.04 | 1.04 | 1.04 | 0.98 | 1.01 | 1 | 1.04 | |
| 4 | | 1.06 | 1.04 | 1.01 | 0.99 | 0.99 | 0.99 | 0.92 | 0.95 | 0.98 | 0.99 | |
| 5 | | 1.02 | 0.97 | 0.97 | 0.94 | 0.94 | 0.94 | 0.91 | 0.95 | 0.92 | 0.94 | |
| 6 | | 0.96 | 0.97 | 0.93 | 0.92 | 0.92 | 0.92 | 0.89 | 0.94 | 0.92 | 0.92 | |
| 7 | | 0.96 | 0.97 | 0.93 | 0.93 | 0.93 | 0.93 | 0.93 | 1 | 0.89 | 0.93 | |
| 8 | | 0.94 | 1.01 | 0.97 | 0.95 | 0.95 | 0.95 | 0.92 | 0.99 | 0.93 | 0.95 | |
| 9 | | 0.91 | 0.96 | 0.96 | 0.94 | 0.94 | 0.94 | 0.96 | 1.04 | 0.96 | 0.94 | |
| 10 | | 0.94 | 0.95 | 0.94 | 0.94 | 0.94 | 0.94 | 0.92 | 0.95 | 0.93 | 0.94 | |
| 11 | | 1 | 1.01 | 0.98 | 0.97 | 0.97 | 0.97 | 0.95 | 0.95 | 0.89 | 0.97 | |
| 12 | | 1 | 1.02 | 1.01 | 0.96 | 0.96 | 0.96 | 0.93 | 0.94 | 0.9 | 0.96 | |
| 13 | | 1.05 | 1 | 1 | 0.98 | 0.98 | 0.98 | 0.96 | 0.95 | 0.93 | 0.98 | |
| 14 | | 1.07 | 0.99 | 1.01 | 1 | 1 | 1 | 1 | 0.99 | 0.89 | 1 | |
| 15 | | | | | | | | | | | | |

(b) Expansion factor CSV file format

|   | A | B | C |
|---|---|---|---|
| 1 | C | Gamma | |
| 2 | 8 | 0.25 | |
| 3 | 1 | 0.5 | |
| 4 | 0.125 | 0.25 | |
| 5 | | | |

(c) SVR Model parameter CSV file format

**FIGURE 2 Input CSV File Formats**

**Output File for AADT Estimation**

Once the input files have been prepared and the locations have been input in the software, the "predict AADT" button should be pressed. While the software is running in the background, a progress bar will appear below the "Predict AADT" button that will show the progress of the software (Figure 3(a)).

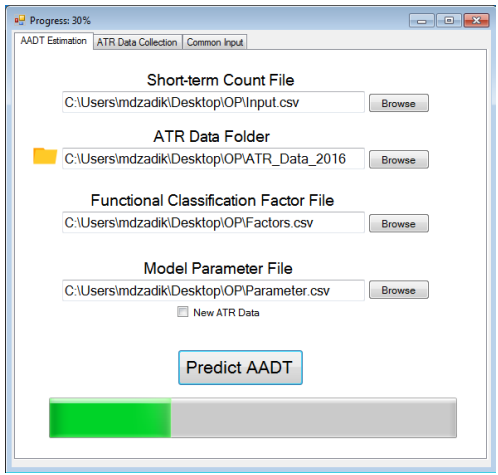

(a) AADT estimation in progress

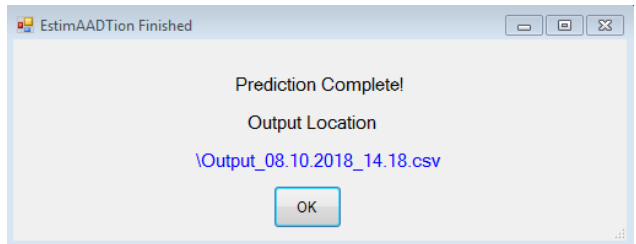

(b) AADT estimation completion

**FIGURE 3 AADT estimation snapshots**

In addition, the progress percentage will appear on the top bar. If the "New ATR" box is unchecked, the software takes about 2-5 minutes to run. However, when "New ATR" box is checked, it can take 3-6 hours for the software to run because the software recreates the models and stores the parameters of the model in the CSV file. During this time, the user can leave the software running but cannot turn off the computer.

After the software run is complete, the progress bar will be full, the runtime menu will disappear and the output menu window will appear (Figure 3(b)). This pop-up window includes a hyperlinked file path to the output file. Clicking on the link will open the output file generated by the software. The output CSV file can also be found in the same folder as the input CSV file. To recognize the output CSV files, each output CSV has a base filename of "Output" appended by the date and time at which it was created. A sample output filename can be "Output_07.18.2018_14.45.CSV." This file was created on 18th July of 2018, at 2:45 PM.

The output CSV file is also of specific format (Figure 4). The output CSV file will always contain five columns. Row 1 contains the labels. Each entry in the output file corresponds to a unique short-term count. If there were multiple entries for one ATR station in the input file, the software aggregates the outputs and provides one output. Column A, B, and C are unchanged from the input file. Column D and E are the new columns which have been generated by the software. Column D is the calculated AADT using the SVR method, and Column E is the calculated AADT from the traditionally used factor-based method. The output from the factor-based method has been included so that users can have a reference for comparison.

| | A | B | C | D | E | F |
|---|---|---|---|---|---|---|
| 1 | County | Station | Functional_Class | AADT-SVR | AADT-Factor | |
| 2 | 1 | 80 | 12 | 12120 | 12140 | |
| 3 | 1 | 99 | 12 | 60454 | 57415 | |
| 4 | 1 | 9 | 2 | 4131 | 3679 | |
| 5 | 1 | 10 | 4 | 2701 | 2493 | |
| 6 | 1 | 11 | 2 | 24432 | 23651 | |
| 7 | 1 | 17 | 4 | 1471 | 1311 | |
| 8 | 1 | 65 | 4 | 1022 | 896 | |
| 9 | 1 | 73 | 4 | 3950 | 3763 | |
| 10 | 1 | 133 | 2 | 2810 | 2678 | |
| 11 | 1 | 149 | 2 | 29021 | 27839 | |
| 12 | 1 | 151 | 4 | 3523 | 3389 | |
| 13 | 1 | 152 | 4 | 4467 | 4274 | |
| 14 | 1 | 153 | 4 | 3612 | 3389 | |
| 15 | 1 | 155 | 4 | 12257 | 11647 | |
| 16 | 1 | 156 | 4 | 2566 | 2333 | |
| 17 | | | | | | |

**FIGURE 4 Output CSV File**

## ANALYSIS AND RESULTS

### AADT estimation model comparison

Before software development, the authors performed a comparative analysis of SVR, ANN and OLS regression method. The authors found that SVR model performs better than the other models for the short term counts in South Carolina. To evaluate the performance of the developed models, pilot tests were conducted in 2016 and 2017 to create an entirely new set of data that were not used in calibration of the models. For this purpose, data were collected from a total of 20 sites that are located throughout the state. Of the 20 ATRs, 3 are located on interstates, 11 are located on arterials, and 6 are located on local roads. From figure 5, it can be observed that the SVR model has 1.80 times less MAPE on average compared to ANN models and 6.5 times less MAPE on average compared to OLS regression model.

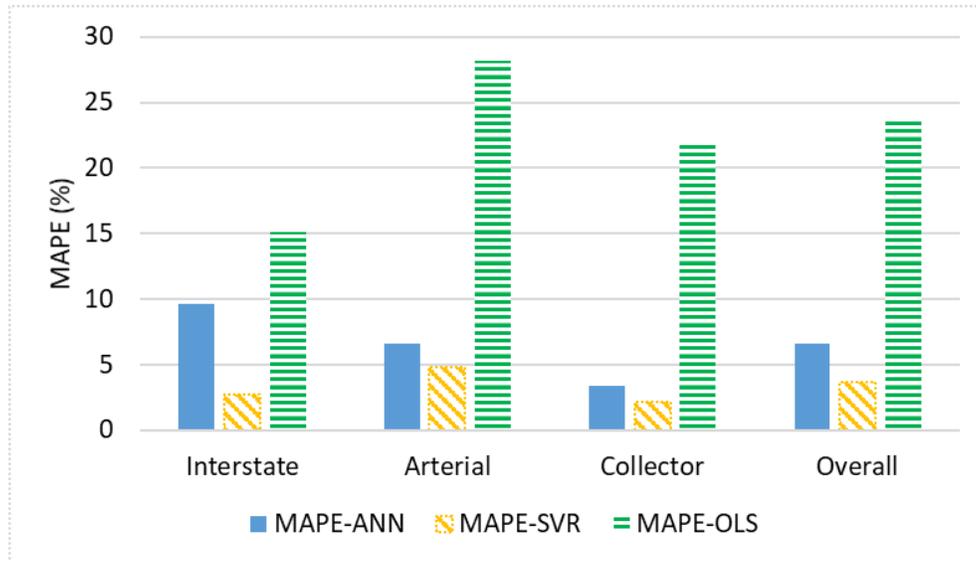

**Figure 5 Comparison of MAPE from ANN, SVR and OLS method for different road classification**

## estimAADTion SOFTWARE EVALUATION

In this study, the estimAADTion software is evaluated with short-term count data collected by SCDOT in different roadway functional classes of SC. In order to compare the efficacy of the software, short-term counts are collected at the locations near the ATR for different roadway functional classes during 2016 and 2017. Then the software is used to predict the AADT. Then the estimated AADT values are compared with the ground truth AADT values (obtained from ATR). Then the MAPE of the software estimation and factor-based estimation are calculated. The results are summarized in Table 1. The results are also aggregated by roadway functional class and shown in Figure 6.

### TABLE 1 Evaluation results of AADT estimation models

| Short-term count station | Nearby ATR | Date | Functional Class | Actual AADT | Factor Method | Software estimation (SVR Method) | MAPE-Factor | MAPE-SVR |
|---|---|---|---|---|---|---|---|---|
| 1 | 80 | 10/19/16 | Interstate | 7,802 | 7,552 | 7,653 | 3% | 2% |
| 2 | 80 | 10/20/16 | Interstate | 7,802 | 8,169 | 7,965 | 5% | 2% |
| 3 | 81 | 10/19/16 | Interstate | 7,341 | 6,855 | 7,052 | 7% | 4% |
| 4 | 81 | 10/20/16 | Interstate | 7,341 | 7,423 | 6,989 | 1% | 5% |
| 5 | 145 | 10/19/16 | Interstate | 21,291 | 21,829 | 22,262 | 3% | 5% |
| 6 | 145 | 10/20/16 | Interstate | 21,291 | 22,337 | 21,231 | 5% | 0% |
| 7 | 100 | 10/19/16 | Interstate | 18,458 | 19,323 | 19,939 | 5% | 8% |
| 8 | 100 | 10/20/16 | Interstate | 18,458 | 19,312 | 18,444 | 5% | 0% |
| 9 | 99 | 10/19/16 | Interstate | 50,518 | 48,233 | 51,251 | 5% | 1% |
| 10 | 99 | 10/20/16 | Interstate | 50,518 | 50,446 | 51674 | 0% | 2% |
| 11 | 9 | 10/24/17 | Arterial | 4,051 | 3,659 | 4,169 | 10% | 3% |

| 12 | 11 | 11/14/17 | Arterial | 24,741 | 22,153 | 23,241 | 10% | 6% |
|----|----|----------|----------|--------|--------|--------|-----|-----|
| 13 | 30 | 11/14/17 | Arterial | 20,707 | 16,181 | 19,623 | 22% | 5% |
| 14 | 38 | 10/31/17 | Arterial | 4,711 | 4,786 | 4,899 | 2% | 4% |
| 15 | 47 | 10/25/17 | Arterial | 24,472 | 21,700 | 24,124 | 11% | 1% |
| 16 | 57 | 11/14/17 | Arterial | 3,274 | 3,124 | 3,418 | 5% | 4% |
| 17 | 133 | 11/7/17 | Arterial | 2,591 | 2,559 | 2,692 | 1% | 4% |
| 18 | 148 | 10/31/17 | Arterial | 28,754 | 24,752 | 28,260 | 14% | 2% |
| 19 | 149 | 10/25/17 | Arterial | 27,269 | 25,989 | 27,517 | 5% | 1% |
| 20 | 17 | 11/7/17 | Collector | 1,352 | 1,259 | 1,357 | 7% | 0% |
| 21 | 73 | 10/24/17 | Collector | 3,555 | 3,817 | 3,750 | 7% | 5% |
| 22 | 152 | 11/7/17 | Collector | 4,203 | 4,096 | 4,192 | 3% | 0% |
| 23 | 153 | 11/7/17 | Collector | 3,437 | 3,427 | 3,407 | 0% | 1% |
| 24 | 154 | 11/7/17 | Collector | 6,953 | 6,162 | 6,840 | 11% | 2% |
| 25 | 155 | 11/7/17 | Collector | 10,784 | 11,079 | 11,215 | 3% | 4% |
| | | | | | | Average | 6% | 3% |

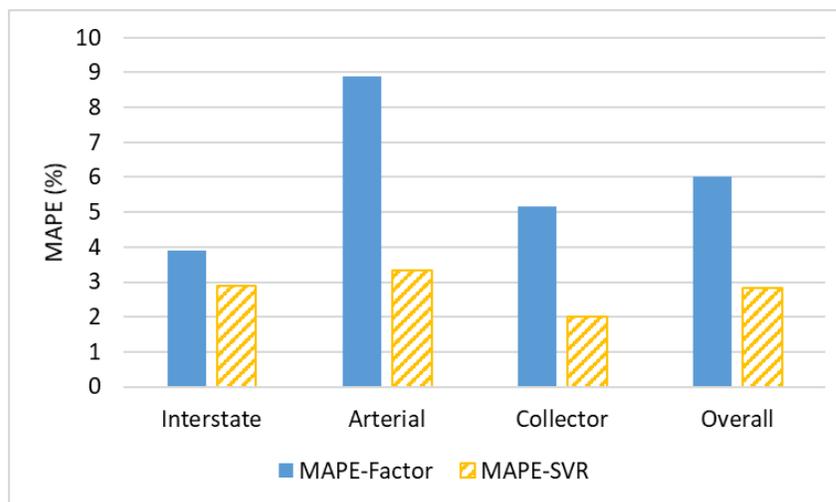

**Figure 6 Comparison of MAPE from factor method and software estimation (SVR method) for different road classification**

## CONCLUSIONS

AADT is an important parameter which is used in numerous transportation projects. As revealed in our study, the traditional practice of AADT estimation using the factor-based method leads to less accurate AADT estimation. However, It was observed from a study by the authors that 97% of the 39 agencies use factor-based AADT estimation model. In this study, a machine learning-based AADT estimation software 'estimAADTion' is developed to enable public agencies with more accurate AADT estimation. This software uses the SVR method to predict AADT from the short-term counts. The estimAADTion software can be used to extract ATR data from any online repository. An example of the website is given in the section of the paper titled "estimAADTion software functions and inputs". Public agencies can use the software to download all ATR data from the previous year. The software is open-source, and it can be modified according to each agency requirements. Also, using historic short-term counts, AADT can be estimated for any specific year. The software also aggregates the results from multiple 24 hour counts and provides only one AADT output for each ATR. Based on the evaluation conducted in this study with the data collected from South Carolina, the MAPE value of the software-estimated AADT is 3%, while the factor-based method produced MAPE of 6%.

The estimAADTion software requires the most updated ATR data to train the model. Therefore, before using the software for AADT estimation, public agencies should run the software in "ATR data collection" mode in order to get the most recent ATR data. It is also recommended to update the parameters of the SVR model whenever there are new ATR data.

## ACKNOWLEDGEMENT

The authors acknowledge the South Carolina Department of Transportation, which provided funding for this research.

## DISCLAIMER

The contents of this report reflect the views of the authors who are responsible for the facts and the accuracy of the presented data. The contents do not reflect the official views of SCDOT or FHWA. This report does not constitute a standard, specification, or regulation.

## AUTHOR CONTRIBUTION STATEMENT

The authors confirm the contribution to the paper as follows: (1) Study conception and design: Zadid Khan, Sakib Mahmud Khan, Mashrur Chowdhury; (2) Data collection: Zadid Khan, Sakib Mahmud Khan; (3) Analysis and interpretation of results: Zadid Khan, Sakib Mahmud Khan, Mashrur Chowdhury; (4) Draft manuscript preparation: Zadid Khan, Sakib Mahmud Khan, Mashrur Chowdhury. All authors reviewed the results and approved the final version of the manuscript.

## REFERENCES

Castro-Neto, M. *et al.* (2009) 'AADT prediction using support vector regression with data-dependent parameters', *Expert Systems with Applications*. doi: 10.1016/j.eswa.2008.01.073.

Chang, C.-C. and Lin, C.-J. (2011) 'LIBSVM', *ACM Transactions on Intelligent Systems and Technology*. doi: 10.1145/1961189.1961199.

Chowdhury, M. *et al.* (2019) *Cost Effective Strategies for Estimating Statewide AADT*. Available at: https://www.scdot.scltap.org/wp-content/uploads/2019/04/SPR-717-Final-Report.pdf (Accessed: 28 July 2019).

Garber, N. J. and Hoel, L. A. (2004) *Traffic and highway engineering*. 5th edn. Cengage Learning.

Islam, S. (2016) *Estimation of Annual Average Daily Traffic (AADT) and Missing Hourly Volume Using Artificial Intelligence*. Clemson Univeristy. Available at: https://tigerprints.clemson.edu/all_theses/2562 (Accessed: 28 July 2019).

Khan, S. M. *et al.* (2018) 'Development of Statewide Annual Average Daily Traffic Estimation Model from Short-Term Counts: A Comparative Study for South Carolina', *Transportation Research Record*. doi: 10.1177/0361198118798979.

Lin, W.-C. (2004) *Case Study on Support Vector Machine Versus Artificial Neural Networks*. University of Pittsburgh. Available at: http://d-scholarship.pitt.edu/8533/.

Mohamad, D. *et al.* (1998) 'Annual Average Daily Traffic Prediction Model for County Roads', *Transportation Research Record: Journal of the Transportation Research Board*. doi: 10.3141/1617-10.

Sharma, S. *et al.* (2001) 'Application of Neural Networks to Estimate AADT on Low-Volume Roads', *Journal of Transportation Engineering*. doi: 10.1061/(asce)0733-947x(2001)127:5(426).

Sharma, S. C. *et al.* (1999) 'Neural Networks as Alternative to Traditional Factor Approach of Annual Average Daily Traffic Estimation from Traffic Counts', *Transportation Research Record: Journal of the Transportation Research Board*, 1660(1), pp. 24–31. doi: 10.3141/1660-04.